\documentclass[11pt, a4paper]{article}
\usepackage{amsmath}

\newtheorem{theorem}{Theorem}

\begin{document}

\title{The Haantjes tensor and double waves for multi-dimensional  systems of hydrodynamic type: a necessary condition for integrability}
\author{E.V. Ferapontov and K.R. Khusnutdinova}
   \date{}
   \maketitle
   \vspace{-7mm}
\begin{center}
Department of Mathematical Sciences \\ Loughborough University \\
Loughborough, Leicestershire LE11 3TU \\ United Kingdom \\[2ex]
e-mails: \\[1ex] \texttt{E.V.Ferapontov@lboro.ac.uk}\\
\texttt{K.Khusnutdinova@lboro.ac.uk}
\end{center}

\bigskip

\begin{abstract}
An  invariant differential-geometric approach to the integrability of $(2+1)$-dimensional
systems of hydrodynamic type,
$$
{\bf u}_t+A({\bf u}) {\bf u}_x+B({\bf u}) {\bf u}_y=0,
$$
is developed. We prove that the existence of special solutions  known as `double waves' is equivalent to the diagonalizability of an arbitrary matrix of the two-parameter family
$$
(kE+A)^{-1}(lE+B).
$$
Since the diagonalizability   can be effectively verified by calculating the Haantjes tensor, this provides a simple  necessary condition for  integrability.


\end{abstract}

\newpage

\section{Introduction}

Over the last 20 years there has been a considerable progress in the theory of  one-dimensional systems of hydrodynamic type ${\bf u}_t+v({\bf u}){\bf u}_x=0$ or, in components, 
\begin{equation}
u^i_t+v^i_j({\bf u}) u^j_x=0, ~~~ i, j=1,..., m,
\label{1d}
\end{equation}
(the standard summation convention over  repeated indices is adopted). Such systems naturally occur in applications in gas dynamics, fluid mechanics, chemical kinetics, Whitham averaging procedure, differential geometry and topological field theory. We refer to \cite{Tsarev, Dub, Serre1, Serre2, Sevennec, Dubr} for a further discussion and references. It has been observed that many particularly important examples are diagonalizable, that is, reducible to the Riemann invariant form
\begin{equation}
R^i_t+v^i({\bf R}) R^i_x=0
\label{1R}
\end{equation} 
where the characteristic speeds $v^i({\bf R})$ satisfy the so-called semi-Hamiltonian property
\cite{Tsarev} (also known as the `richness' condition \cite{Serre2}), 
\begin{equation}
\partial_k\left(\frac{\partial_j v^i}{v^j-v^i} \right)= \partial_j\left(\frac{\partial_k v^i}{v^k-v^i} \right), 
\label{semi}
\end{equation}
$\partial_k=\partial/\partial R^k$, \  $i\ne j\ne k$. We emphasize that the semi-Hamiltonian property (\ref{semi}) is usually automatically satisfied for  diagonalizable systems of the `physical' origin. For instance, a conservative diagonalizable system is necessarily semi-Hamiltonian. We recall that an $n$-component system (\ref{1d}) is said to be conservative if it possesses $n$ conservation laws of hydrodynamic type whose densities are functionally independent. It turns out that the additional requirement of the diagonalizability  implies the existence of an infinity of conservation laws and, hence, the semi-Hamiltonian property: see \cite{Sevennec}, or the Appendix to \cite{Fer7} for a simpler proof. Semi-Hamiltonian systems  possess infinitely many
conservation laws and commuting flows of hydrodynamic type and can be linearized by the generalized hodograph method \cite{Tsarev}.  Their analytic, differential-geometric and Hamiltonian aspects are 
well-understood by now. 

Remarkably, there exist the effective tensor criteria to verify the diagonalizability and the semi-Hamiltonian property  without the actual computation of  eigenvalues and  eigenvectors of the matrix $v^i_j$.
Let us first calculate the Nijenhuis tensor of the matrix $v^i_j$,
\begin{equation}
N^i_{jk}=v^p_j\partial_{u^p}v^i_k-v^p_k\partial_{u^p}v^i_j-v^i_p(\partial_{u^j}v^p_k-\partial_{u^k}v^p_j),
\label{N}
\end{equation}
 and introduce the Haantjes tensor
\begin{equation}
H^i_{jk}=N^i_{pr}v^p_jv^r_k-N^p_{jr}v^i_pv^r_k-N^p_{rk}v^i_pv^r_j+N^p_{jk}v^i_rv^r_p.
\label{H}
\end{equation}
For strictly hyperbolic systems  the condition of diagonalizablity is given by the following theorem
(which was stated in \cite{Haantjes} in  purely geometric terms as a condition of diagonalizability of a (1,1) tensor field).
\begin{theorem} \cite{Haantjes}
 A  hydrodynamic type system (\ref{1d}) with  mutually distinct characteristic speeds is diagonalizable if and only if the corresponding Haantjes tensor (\ref{H}) is identically zero. 
\end{theorem}
Since  components of the Haantjes tensor can be obtained using any  computer algebra package, this provides an effective diagonalizability criterion. This criterion has been successfully implemented in \cite{Fer3} to classify isotherms of adsorption for which the  equations of chromatography possess Riemann invariants (since the equations of chromatography are conservative, the semi-Hamiltonian property will be automatically satisfied). The same criterion was applied in \cite{Tsarev1} to the Whitham equations governing slow modulations of  traveling waves for the generalized KdV equation $u_t+f(u)u_x+u_{xxx}=0$. It was demonstrated that the 
Whitham equations are diagonalizable (and hence semi-Hamiltonian due to their conservative nature)
if and only if $f'''=0$.
Although, for conservative systems,  the diagonalizability implies the semi-Hamiltonian property, this is not true in general. The tensor object responsible for the semi-Hamiltonian property was  introduced in \cite{Pavlov2} (see the Appendix).

The present paper aims at the discussion of the extent to which the one-dimensional theory carries over to $(2+1)$-dimensional quasilinear systems
\begin{equation}
{\bf u}_t+A({\bf u}) {\bf u}_x+B({\bf u}) {\bf u}_y=0;
\label{1}
\end{equation}
here  ${\bf u}$ is an
$m$-component column vector and $A({\bf u}), B({\bf u})$ are $m\times
m$ matrices. Systems of this type describe many physical phenomena.
In particular, important examples occur in gas dynamics, shallow
water theory, combustion theory, general relativity, nonlinear elasticity, magneto-fluid
dynamics, etc \cite{Majda, Dafermos}. Particularly  interesting `integrable' systems of the form (\ref{1}) arise as dispersionless limits of multi-dimensional
 soliton equations \cite{Zakharov}, within the method of Whitham averaging applied to `integrable' two-dimensional models \cite{Kr1, Kr2}, and  the R-matrix approach
\cite{Bla}.

The first natural restriction to impose is that all systems arising as  one-dimensional limits of (\ref{1}), 
in particular, the systems ${\bf u}_t+A({\bf u}) {\bf u}_x=0$ and ${\bf u}_t+B({\bf u}) {\bf u}_y=0$,  are diagonalizable (but not simultaneously, as the matrices  $A$ and $B$ do not commute in general). Furthermore, applying to (\ref{1}) an arbitrary linear change of the independent variables,
$$
\tilde t= a_{11}t+a_{12}x+a_{13}y, ~~ \tilde x= a_{21}t+a_{22}x+a_{23}y, ~~
\tilde y= a_{31}t+a_{32}x+a_{33}y,
$$
we arrive at the transformed system
$$
{\bf u}_{\tilde t}+\tilde A({\bf u}) {\bf u}_{\tilde x}+\tilde B({\bf u}) {\bf u}_{\tilde y}=0
$$
where
$$
\begin{array}{c}
\tilde A=(a_{11}E+a_{12}A+a_{13}B)^{-1}(a_{21}E+a_{22}A+a_{23}B), \\
\tilde B=(a_{11}E+a_{12}A+a_{13}B)^{-1}(a_{31}E+a_{32}A+a_{33}B),
\end{array}
$$
and $E$ is the $m\times m$ identity matrix. Since we want our approach to be invariant under linear changes of variables, we  require that all matrices of the multi-parameter family
\begin{equation}
(aE+bA+cB)^{-1}(\tilde aE+\tilde bA+\tilde cB)
\label{fam}
\end{equation}
are diagonalizable and semi-Hamiltonian.  These considerations motivate the following definitions:

\medskip

\noindent {\bf Definition 1} {\it A $(2+1)$-dimensional system  is said to be {\it diagonalizable} if an arbitrary matrix of the family
(\ref{fam}) is diagonalizable.}

\medskip

\noindent {\bf Definition 2} {\it A diagonalizable $(2+1)$-dimensional system  is said to be {\it semi-Hamiltonian}  if an arbitrary matrix of the family (\ref{fam}) is  semi-Hamiltonian.}

\medskip

\noindent {\bf Remarks.} One can show that some  parameters in (\ref{fam}) are, in fact, redundant: it is  sufficient to verify the diagonalizability and the semi-Hamiltonian property for an arbitrary matrix in the smaller family
\begin{equation}
(kE+A)^{-1}(lE+B)
\label{small}
\end{equation}
where $k$ and $l$ are arbitrary constants. Indeed, one can simplify the general matrix (\ref{fam}) using the fact that the inversion and the addition of  a  multiple of the identity do not effect the diagonalizability. 

We point out that for many systems (\ref{1}) the diagonalizability is already sufficiently restrictive and implies the semi-Hamitonian property. This is the case, for instance, if the original two-dimensional system (\ref{1}) is conservative. Indeed, all one-dimensional limits of a multi-dimensional conservative system  inherit the  conservative form, and in (1+1) dimensions the diagonalizability is known to imply the semi-Hamiltonian property. 

Finally, we point out that both definitions generalize to  multi-dimensional setting (3+1, etc)  in an obvious way. 

Examples of diagonalizable semi-Hamiltonian three-component systems (\ref{1}) are discussed in Sect. 2. The classification of a special class of diagonalizable three-component conservative systems is obtained in Sect. 4.

\bigskip

An alternative approach to the integrability of multi-dimensional systems of hydrodynamic type is based on the method of hydrodynamic reductions. The key element of this construction are  exact solutions of the
system (\ref{1}) of the form
${\bf u(\bf R)}={\bf u}(R^1, ..., R^n)$ where the  Riemann invariants
$R^1, ..., R^n$ solve a pair of
commuting diagonal systems
\begin{equation}
R^i_t=\lambda^i({\bf R })\ R^i_x, ~~~~ R^i_y=\mu^i({\bf R })\ R^i_x;
\label{R}
\end{equation}
notice that the number of Riemann invariants is allowed to be
arbitrary.  Thus, the original  $(2+1)$-dimensional system (\ref{1}) is
decoupled into a compatible pair of diagonal $(1+1)$-dimensional systems.
Solutions of this type are known as multiple waves, or nonlinear interactions of $n$ planar
simple waves (also called  solutions with a degenerate hodograph: simple waves (double waves) when the number of Riemann invariants equals one (two)). These solutions were extensively investigated in gas dynamics and
magnetohydrodynamics in a series of publications \cite{Sidorov, Burnat1,
Burnat2, Burnat3, Perad1, Perad2, Dinu, Grundland}. Later, they 
reappeared in the context of the dispersionless KP and Toda hierarchies
\cite{ Gibb94, GibTsa96, GibTsa99, GuMaAl,  Ma, Ma1, Lei, Fer},
the theory of  integrable hydrodynamic chains \cite{Pavlov,
Pavlov1, Shabat} and the Laplacian growth problems \cite{Kr3}. In \cite{Fer4}, it was suggested to call a
multi-dimensional system {\it integrable} if, for arbitrary $n$,  it possesses
infinitely many $n$-component reductions of the form (\ref{R})
 parametrized by $n$ arbitrary functions of a single argument.  It was shown that this requirement provides an effective classification criterion. Partial classification results were obtained in \cite{Fer4, Fer5, Fer6}. It was demonstrated in \cite{Fer2, Fer6} that the method of hydrodynamic reductions is effective in any dimension: in particular, $(3+1)$- and $(5+1)$-dimensional integrable examples were uncovered. 
We recall, see \cite{Tsarev}, that the requirement of  the
commutativity of the flows (\ref{R})
is equivalent to the following restrictions on their characteristic speeds:
\begin{equation}
\frac{\partial_j\lambda
^i}{\lambda^j-\lambda^i}=\frac{\partial_j\mu^i}{\mu^j-\mu^i},
\label{comm}
\end{equation}
$i\ne j,
~ \partial_j=\partial/\partial_{ R^j}$, no summation! Once these conditions are met, the general solution
of the system (\ref{R}) is given by the
implicit  `generalized hodograph'  formula \cite{Tsarev},
\begin{equation}
v^i({\bf R })=x+\lambda^i({\bf R })\ t+\mu^i({\bf R }) \ y, 
\label{hod}
\end{equation}
$i=1, ..., n$. Here $v^i({\bf R })$ are  the characteristic speeds of the general flow
commuting with (\ref{R}), that is, the general solution of the linear
system
\begin{equation}
\frac{\partial_jv^i}{v^j-v^i}=\frac{\partial_j\lambda
^i}{\lambda^j-\lambda^i}=\frac{\partial_j\mu^i}{\mu^j-\mu^i}.
\label{comm1}
\end{equation}
Substituting ${\bf u}(R^1, ..., R^n)$ into (\ref{1}) and using
(\ref{R}) one  arrives at the equations
\begin{equation}
(A+\mu^iB+\lambda^i E)\ \partial_i{\bf u}=0, 
\label{2}
\end{equation}
(no summation)  implying that both $\lambda^i$ and $\mu^i$ satisfy the dispersion relation
\begin{equation}
D(\mu, \lambda)={\rm det} (A+\mu B+\lambda E)=0.
\label{dispersion}
\end{equation}
Thus, the construction of solutions describing nonlinear interactions of $n$ planar simple
waves consists of two steps:

\noindent {\bf (1)} Reduce the initial system (\ref{1}) to a pair of
commuting flows (\ref{R}) by solving the equations  (\ref{comm}),
(\ref{2}) for ${\bf u}({\bf R }), \ \lambda^i({\bf R }), \ \mu^i({\bf R })$ as functions
of the Riemann invariants $R^1, ..., R^n$. These
equations are highly overdetermined and
do not possess solutions in general. However, for {\it integrable} systems these reductions depend, modulo reparametrizations $R^i\to f^i(R^i)$,  on $n$ arbitrary functions of  a single argument. Once a particular
reduction of the form (\ref{R}) is constructed, the second step is
fairly straightforward:

\noindent {\bf (2)} Solve the linear system (\ref{comm1}) for
$v^i({\bf R })$ and determine $R^1, ..., R^n$ as functions of $t, x, y$ from
the implicit  hodograph formula (\ref{hod}). This step provides some extra $n$ arbitrary functions. 

\noindent Therefore, solutions arising within this scheme depend on $2n$ essential functions of a single argument.

\medskip

\noindent {\bf Simple waves.} For $n=1$ we have
${\bf u}={\bf u}(R)$, thus, hodographs of these solutions are curves. The scalar variable $R=R^1$ solves a
pair of first order PDEs
$$
R_t=\lambda (R)\ R_x, ~~~~ R_y=\mu (R)\ R_x
$$
which, in the  one-component situation, are automatically commuting. The  hodograph formula (\ref{hod})
takes the  form
\begin{equation}
f(R)=x+\lambda(R)t+\mu(R)y
\label{P1}
\end{equation}
where $f(R)$ is  arbitrary. This formula  shows that, in coordinates
$t, x, y$,  the surfaces $R=const$ are planes so that the solution
${\bf{u}}={\bf u}(R)$ is constant along a one-parameter family of
planes. Therefore,  it is singular along the developable surface which envelopes this one-parameter family. Solutions of this type, known as planar simple waves, exist
for all multi-dimensional quasilinear systems  and do not impose any restrictions on the  matrices $A$ and $B$. 

\medskip

\noindent {\bf Double waves.} For  $n=2$ we have
${\bf u}={\bf u}(R^1, R^2)$ so that hodographs of these solutions (known as double waves, or nonlinear interactions of two planar simple waves) are surfaces. 
Double waves, as well as simple waves, belong to the class of solutions with a `degenerate hodograph'. In the context of multi-dimensional gas dynamics, they were extensively investigated in \cite{Sidorov}.  Notice that each hodograph surface corresponds to infinitely many solutions, indeed, one needs to solve  the system (\ref{R}) for 
$R^1, R^2$ to make a hodograph surface into a solution. This step does not change the hodograph surface, it just specifies the dependence on $t, x, y$. The general solution of (\ref{R}) is
given by the  implicit hodograph formula
\begin{equation}
v^1({\bf R })=x+\lambda^1({\bf R })t+\mu^1({\bf R })y, ~~~
v^2({\bf R })=x+\lambda^2({\bf R })t+\mu^2({\bf R })y.
\label{P2}
\end{equation}
Setting $R^1=const, ~ R^2=const$, one obtains a two-parameter family
of lines (or, using  differential-geometric language, a line congruence) in the
3-space of
independent variables $t, x, y$. The corresponding solution ${\bf
u}={\bf u}(R^1, R^2)$ is constant along the lines of this two-parameter
family. Therefore,  it is singular along the  focal surfaces of the congruence.

One can show that any $2\times 2$  system (\ref{1}) possesses
infinitely many two-component reductions of the form (\ref{R})
parametrized by two arbitrary functions of a single  argument. The corresponding double waves depend on four arbitrary functions. On the contrary, for multi-component systems  $(m\geq3)$ the requirement of  the existence of double waves imposes strong restrictions on the matrices $A$ and $B$. In Sect. 3 we prove our main result

\begin {theorem}  The Haantjes tensor of an arbitrary matrix (\ref{small}) is zero if and only if 

\noindent (i) the system (\ref{1})  possesses double waves
parametrized by four arbitrary functions of a single argument;

\noindent (ii) the characteristic speeds $\lambda^i, \ \mu^i$ of two-component reductions are not restricted by any  algebraic relations other than  the dispersion relation $D(\mu, \lambda)=0$. That is, for any point ${\bf u}_0$ in the hodograph space and any two points $(\lambda^1, \mu^1)$ and $(\lambda^2, \mu^2)$ on the dispersion curve at  ${\bf u}_0$ one can find a two-component reduction whose characteristic speeds at this point take the values $(\lambda^1, \lambda^2)$ and $(\mu^1, \mu^2)$, respectively.
\end{theorem}

The condition  (ii)  is crucial: for instance, the equations of  two-dimensional 
gas dynamics possess potential double waves parametrized by four arbitrary functions, however, the system is  not diagonalizable, see Sect. 5. The explanation of this phenomenon  lies in the fact that  the characteristic speeds of commuting flows defining two-component reductions come from one and the same branch of the dispersion curve (which is reducible), so that the condition (ii) is violated.


Thus, we have  an easy-to-verify necessary condition for the integrability of multi-component multi-dimensional systems of hydrodynamic type. It should be emphasized that the condition of diagonalizability is necessary, but not at all sufficient for the integrability. In Sect. 2 we construct an example of a diagonalizable semi-Hamiltonian $3\times 3$ system in $2+1$ dimensions which is not integrable for generic values of parameters  (Example 2). 


\medskip

\noindent {\bf Multiple waves.} The requirement of  the existence of nontrivial
3-component reductions imposes further constraints on $A$ and $B$. These prove to be very
restrictive and imply the
existence of  $n$-component reductions for arbitrary $n$ \cite{Fer4}. 
This  phenomenon is similar to the well-known three-soliton condition in
the Hirota bilinear approach which, generically, implies the existence of n-soliton solutions and the integrability. For two-component systems (\ref{1}) the full set of constraints imposed on $A$ and $B$ by the requirement of   existence of three-component reductions was obtained in \cite{Fer5}. These constraints imply, in particular, that all 2-component integrable systems of the form (\ref{1}) are necessarily symmetrizable in the sense of \cite{Godunov} (that is, possess three conservation laws of hydrodynamic type). Moreover, they necessarily possess a scalar pseudopotential playing the role of a dispersionless `Lax pair'. We expect that both  results  generalize to a multi-component situation.

We emphasize that, although the method of hydrodynamic reductions provides infinitely many (implicit) solutions parametrised by arbitrarily many functions of a single argument, 
the question of solving the initial value problem for integrable systems (\ref{1}) remains open. We believe, however, that solutions describing nonlinear interactions of planar simple waves are locally dense in the space of all solutions of (\ref{1}) (see \cite{GibTsa99} for a discussion of this issue in the context of the dispersionless KP equation). A detailed investigation of their breakdown and singularity structure is important for the analysis of the general Cauchy problem for multi-dimensional quasilinear systems. We hope that the combination of the available existence results (for the Cauchy problem for semi-Hamiltonian systems in the dimension one) with the decomposition  of an integrable multi-dimensional system into a collection of commuting semi-Hamiltonian  flows would lead to a new understanding of the Cauchy problem in many dimensions.

\section{Examples}

In this section we list some examples of diagonalizable semi-Hamiltonian $3\times 3$ systems of hydrodynamic type in $2+1$ dimensions.

{\bf Example 1.}  Let us consider the so-called  generalized Benney system \cite{Zakharov},
$$
a_t+(av)_x=0, ~~~ v_t+vv_x+w_x=0, ~~~ w_y+a_x=0,
$$
 which reduces to the shallow water equations in the limit $y=-x, \ w=a$. In matrix form, we have
 ${\bf u}_x+A({\bf u}) {\bf u}_t+B({\bf u}) {\bf u}_y=0$, here ${\bf u}=(a, v, w)^{t}$ or, explicitly, 
$$
\left(\begin{array}{c}
a \\
v\\
w
\end{array}
\right)_x+
\left(\begin{array}{ccc}
0 & 0 & 0\\
1/a & 0 & 0\\
-v/a&1&0
\end{array}
\right)
\left(\begin{array}{c}
a \\
v\\
w
\end{array}
\right)_t+
\left(\begin{array}{ccc}
0 & 0 & 1\\
0 & 0 & -v/a\\
0&0&v^2/a
\end{array}
\right)
\left(\begin{array}{c}
a \\
v\\
w
\end{array}
\right)_y=0.
$$
We have verified that the Haantjes tensor of the corresponding matrix (\ref{small}) is zero. Since the  system possesses four conservation laws
$$
\begin{array}{c}
a_t+(av)_x=0,  ~~~ v_t+(v^2/2+w)_x=0, \\
\ \\
w_y+a_x=0, ~~~ (av)_t+(aw+av^2)_x+(w^2/2)_y=0,
\end{array}
$$
the semi-Hamiltonian property is automatically satisfied (we have  verified it independently using the tensor criterion from the Appendix). 

Looking for hydrodynamic reductions of the generalized Benney system in the form $a=a(R^1, ..., R^n)$, $ b=b(R^1, ..., R^n)$,  $ w=w(R^1, ..., R^n)$,
where the Riemann invariants $R^i$ satisfy  (\ref{R}), one  obtains
\begin{equation}
\partial_i w=-(\lambda^i+v)  \partial_i v, ~~~ \partial_i a=\mu^i
(\lambda^i+v)  \partial_i v,
\label{60}
\end{equation}
along with the dispersion relation
\begin{equation}
 \mu^i=-\frac{a}{(\lambda^i+v)^2}.
\label{59}
\end{equation}
The compatibility condition
$\partial_i\partial_jw=\partial_j\partial_iw$ implies
\begin{equation}
\partial_i\partial_jv=\frac{\partial_j\lambda^i}{\lambda^j-\lambda^i}\partial_iv+\frac{\partial_i\lambda^j}{\lambda^i-\lambda^j}\partial_jv,
\label{61}
\end{equation}
while the commutativity condition (\ref{comm}) results in
\begin{equation}
\partial_j\lambda^i=\frac{\lambda^j+v}{\lambda^i-\lambda^j}\partial_j v.
\label{62}
\end{equation}
The substitution of (\ref{62}) into (\ref{61}) implies the   system
for $v(R)$ and $\lambda^i(R)$,
\begin{equation}
\partial_j\lambda^i=\frac{\lambda^j+v}{\lambda^i-\lambda^j}\partial_j v, ~~~
\partial_i\partial_jv=-\frac{\lambda^i+\lambda^j+2v}{(\lambda^j-\lambda^i)^2}\partial_iv\partial_jv.
\label{63}
\end{equation}
One can verify that the remaining compatibility conditions
$\partial_i\partial_ja=\partial_j\partial_ia$ are satisfied
identically.
For any solution $\lambda^i,  v$ of the system (\ref{63}) one can
reconstruct $w, a, \mu^i$ by  virtue of  (\ref{60}), (\ref{59}). The system
(\ref{63}) is compatible, with the general solution  depending on $n$
arbitrary functions of a single argument (modulo  reparametrizations $R^i\to
f^i(R^i)$), thus manifesting the integrability of the generalized Benney system.  We also recall that the  generalized Benney system arises as a consistency condition of the dispersionless Lax pair \cite{Zakharov}, 
$$
\psi_t=-\frac{1}{2}\psi_x^2-w, ~~~ \psi_y=\frac{a}{\psi_x-v}.
$$
Two-component reductions  are governed by the equations
$$
\partial_2\lambda^1=\frac{\lambda^2+v}{\lambda^1-\lambda^2}\partial_2 v, ~~~ \partial_1\lambda^2=\frac{\lambda^1+v}{\lambda^2-\lambda^1}\partial_1 v, ~~~
\partial_1\partial_2v=-\frac{\lambda^1+\lambda^2+2v}{(\lambda^2-\lambda^1)^2}\partial_1v\partial_2v
$$
whose general solution depends, modulo reparametrizations $R^1\to
f^1(R^1), \ R^2\to f^2(R^2)$, on two arbitrary functions of a single variable. Two extra arbitrary functions come from the solution of the corresponding hydrodynamic type systems (\ref{comm}). Since the characteristic speeds $(\lambda^1, \ \mu^1)$ and $(\lambda^2, \ \mu^2)$ are not restricted by any algebraic relations other then the dispersion relation (\ref{59}), both conditions of Theorem 2 are satisfied.

{\bf Example 2.} 
Let us consider a class of symmetric systems (\ref{1}) for which
$A$ is a constant diagonal matrix and $B$ is a Hessian:
\begin{equation}
\left(\begin{array}{c}
u \\
v\\
w
\end{array}
\right)_t=
\left(\begin{array}{ccc}
a & 0 & 0\\
0 & b & 0\\
0&0&c
\end{array}
\right)
\left(\begin{array}{c}
u \\
v\\
w
\end{array}
\right)_x+
\left(\begin{array}{ccc}
F_{uu} & F_{uv} & F_{uw}\\
F_{uv} & F_{vv} & F_{vw}\\
F_{uw}&F_{vw}&F_{ww}
\end{array}
\right)
\left(\begin{array}{c}
u \\
v\\
w
\end{array}
\right)_y.
\label{ex}
\end{equation}
Here $c<b<a$ are constants and the potential $F(u, v, w)$ is of the form
\begin{equation}
F=v(\gamma u+\delta w)+f(\gamma u+\delta w)
\label{Fdeg}
\end{equation} 
where the  constants $\gamma$ and $\delta$ satisfy the constraint
$(b-a)\delta^2+(b-c)\gamma^2=0$ and $f(\cdot)$ is an arbitrary function of a single argument
(one can set $\gamma=1/\sqrt{b-c}, \ \delta=1/\sqrt{a-b}$). Potentials of this type arise  in the classification of diagonalizable systems of the form (\ref{ex}) -- see Sect. 4. We have verified that the system (\ref{ex}), (\ref{Fdeg}) is diagonalizable and semi-Hamiltonian. In fact, the semi-Hamiltonian property follows from the conservative nature of 
(\ref{ex}).  In spite of  
being diagonalizable and semi-Hamiltonian, these systems are not integrable in general (that is, do not  possess $n$-wave solutions for $n\geq3$). The integrability conditions impose
one extra constraint on the potential (\ref{Fdeg}),
$$
f''''=\gamma^2\frac{c-b}{c-a}f'' (2(f''')^2-f''f'''').
$$
This ODE can be  solved explicitly implying that, up to elementary changes of variables,
 \begin{equation}
 f''=\frac{1}{\sqrt \epsilon} \cot (\gamma u+\delta w), ~~~ \epsilon=\gamma^2\frac{c-b}{c-a},
 \label{fdeg}
 \end{equation}
provided $f'''\ne 0$; otherwise the system (\ref{ex}) is linear.  Notice that we need $f''$ (rather then $f$ itself) to write down equations (\ref{ex}):
\begin{equation}
\left(\begin{array}{c}
u \\
v\\
w
\end{array}
\right)_t=
\left(\begin{array}{ccc}
a & 0 & 0\\
0 & b & 0\\
0&0&c
\end{array}
\right)
\left(\begin{array}{c}
u \\
v\\
w
\end{array}
\right)_x+
\left(\begin{array}{ccc}
\gamma^2f'' & \gamma &  \gamma \delta f''\ \\
\gamma & 0 & \delta \\
 \gamma \delta f'' &  \delta  & \delta^2 f''
\end{array}
\right)
\left(\begin{array}{c}
u \\
v\\
w
\end{array}
\right)_y.
\label{K}
\end{equation}
For $f''$ given by (\ref{fdeg}) this system possesses the scalar pseudopotential
$$
\begin{array}{c}
\psi_t=\frac{bv}{\sqrt \epsilon}+\frac{a}{\gamma^2}\log \sin (\gamma u-p\psi_y)+
\frac{c}{\delta^2}\log \sin (\delta w+p\psi_y), \\
\ \\
\psi_x=\frac{v}{\sqrt \epsilon}+\frac{1}{\gamma^2}\log \sin (\gamma u-p\psi_y)+
\frac{1}{\delta^2}\log \sin (\delta w+p\psi_y),
\end{array}
$$
where $p=\frac{\gamma^2\sqrt{\epsilon}}{b-a}$ (that is, the system (\ref{K})  arises from the compatibility condition $\psi_{tx}=\psi_{xt}$).  Thus, this pseudopotential can be viewed as the dispersionless analogue of the Lax pair. In the $2\times 2$ case, the existence of a scalar pseudopotential is the necessary and sufficient condition for the integrability of a multi-dimensional quasilinear system \cite{Fer5}. 
This  should be true in a multi-component situation as well, although the  proof meets  technical difficulties. 

\noindent Notice that the dispersion relation for the system (\ref{K}),
$$
det \left(\lambda E-\mu \left(\begin{array}{ccc}
a & 0 & 0\\
0 & b & 0\\
0&0&c
\end{array}
\right)
-\left(\begin{array}{ccc}
\gamma^2f'' & \gamma &  \gamma \delta f''\ \\
\gamma & 0 & \delta \\
 \gamma \delta f'' &  \delta  & \delta^2 f''
\end{array}
\right)\right)=0,
$$
factorises into a line and a conic,
$$
(\lambda-b\mu)[(\lambda-a\mu)(\lambda-c\mu)-(\gamma^2+\delta^2)(1+f''(\lambda-b\mu))]=0.
$$
Thus, the system (\ref{K})  provides an integrable example from the class discussed in \cite{K}. 

\bigskip

Below we list some further examples of three-component integrable $(2+1)$-dimensional systems of the form
(\ref{1}) which were constructed in \cite{Bla} using the classical $R$-matrix approach (see also \cite{Ma1}). We have verified  that all of them are diagonalizable and semi-Hamiltonian by directly computing the corresponding tensors (\ref{H}) and (\ref{P}). In fact, the semi-Hamiltonian property follows from their conservative nature. For each of these examples we have calculated all conservation laws of hydrodynamic type and  scalar pseudopotentials which play the role of dispersionless Lax pairs.
We hope that these examples will be useful for a further research in the area of multi-dimensional systems of hydrodynamic type.

{\bf Example 3.} The 3-component system
$$
\left(\begin{array}{c}
u \\
v\\
w
\end{array}
\right)_t=
\left(\begin{array}{ccc}
0 & 2 & 0\\
-u & 0 & 2\\
-v/2&0&0
\end{array}
\right)
\left(\begin{array}{c}
u \\
v\\
w
\end{array}
\right)_x+
\left(\begin{array}{ccc}
0 & 0 & 0\\
0 & 0 & 0\\
1/2&0&0
\end{array}
\right)
\left(\begin{array}{c}
u \\
v\\
w
\end{array}
\right)_y
$$
constitutes the `second' commuting flow in the dispersionless KP hierarchy. It possesses  four conservation laws of hydrodynamic type,
$$
\begin{array}{c}
u_t=2v_x, \\
\ \\
v_t=(2w-u^2/2)_x, \\
\ \\
(2w-u^2/4)_t=u_y-(uv)_x, \\
\ \\
(4uw+2v^2-u^3/2)_t=(u^2)_y+(8vw-3u^2v)_x,
\end{array}
$$
and the scalar pseudopotential
$$
\begin{array}{c}
\psi_t=\frac{1}{2}\psi_x^2+u, \\
\ \\
\psi_y=\frac{1}{8}\psi_x^4+\frac{1}{2}u\psi_x^2+v\psi_x+2w.
\end{array}
$$
The system arises from  the compatibility condition $\psi_{ty}=\psi_{yt}$. 

{\bf Example 4.} The 3-component system
$$
\left(\begin{array}{c}
u \\
v\\
w
\end{array}
\right)_t=
\left(\begin{array}{ccc}
0 & 2-r & 0\\
rv & (1-r)u & 2-r\\
(1+r)w&0&(1-r)u
\end{array}
\right)
\left(\begin{array}{c}
u \\
v\\
w
\end{array}
\right)_x+
\left(\begin{array}{ccc}
1 & 0 & 0\\
0 & 0 & 0\\
0&0&0
\end{array}
\right)
\left(\begin{array}{c}
u \\
v\\
w
\end{array}
\right)_y,
$$
$r$=const, possesses four conservation laws
$$
\begin{array}{c}
u_t=u_y+(2-r)v_x, \\
\ \\
\left(w^{\frac{1-r}{1+r}}\right)_t=(1-r)\left(w^{\frac{1-r}{1+r}}u\right)_x, \\
\ \\
(u^2+\frac{2(2-r)}{2r-1}v)_t=(u^2)_y+\frac{2(2-r)}{2r-1}((2-r)w+ruv)_x, \\
\ \\
\left(w^{\frac{1-2r}{1+r}}v    \right)_t=\left( (1-r)uvw^{\frac{1-2r}{1+r}}+(1+r)w^{\frac{2-r}{1+r}}\right)_x.
\end{array}
$$
and the scalar pseudopotential
$$
\begin{array}{c}
\psi_t=(1-r)\left(u\psi_x+\psi_x^{\frac{2-r}{1-r}}\right), \\
\ \\
\psi_y=(1-r)\left(u\psi_x+v\psi_x^{\frac{r}{r-1}}+w\psi_x^{\frac{r+1}{r-1}}+\psi_x^{\frac{2-r}{1-r}}\right).
\end{array}
$$

{\bf Example 5.} The 3-component system
$$
\left(\begin{array}{c}
u \\
v\\
w
\end{array}
\right)_t=
\left(\begin{array}{ccc}
(r-1)v & (2-r)u & 0\\
rw & 0 & (2-r)u\\
1+r&0&0
\end{array}
\right)
\left(\begin{array}{c}
u \\
v\\
w
\end{array}
\right)_x+
\left(\begin{array}{ccc}
1 & 0 & 0\\
0 & 0 & 0\\
0&0&0
\end{array}
\right)
\left(\begin{array}{c}
u \\
v\\
w
\end{array}
\right)_y,
$$
$r$=const, possesses four conservation laws
$$
\begin{array}{c}
w_t=(1+r)u_x, \\
\ \\
\left(u^{\frac{r-1}{2-r}}\right)_t=\left(u^{\frac{r-1}{2-r}}\right)_y+(r-1)\left(u^{\frac{r-1}{2-r}}v\right)_x, \\
\ \\
\left(v+\frac{1-r}{1+r}w^2\right)_t=(2-r)(uw)_x, \\
\ \\
\left(\frac{1+r}{3-2r}u+vw+\frac{2-3r}{6(1+r)}w^3\right)_t=\frac{1+r}{3-2r}u_y+
\left(\frac{2-r}{2}uw^2+\frac{(1+r)(2-r)}{3-2r}uv \right)_x,
\end{array}
$$
and the scalar pseudopotential
$$
\begin{array}{c}
\psi_t=(1-r)u\psi_x^{\frac{2-r}{1-r}}, \\
\ \\
\psi_y=(1-r)\left(u\psi_x^{\frac{2-r}{1-r}}+v\psi_x+w\psi_x^{\frac{r}{r-1}}+\psi_x^{\frac{r+1}{r-1}}\right).
\end{array}
$$

\section{Proof of Theorem 2}

Let us first clarify where the obstruction to the existence of double waves comes from. To construct double waves one needs to solve the commutativity equations (\ref{comm}),
$$
\frac{\partial_j\lambda
^i}{\lambda^j-\lambda^i}=\frac{\partial_j\mu^i}{\mu^j-\mu^i},
$$
 along with the relations (\ref{2}), 
$$
(A+\mu^iB+\lambda^i E)\ \partial_i{\bf u}=0,
$$
here $i, j =1, 2$ and ${\bf u}={\bf u}(R^1, R^2)$. The last condition implies the dispersion relation
$$
det (A+\mu^iB+\lambda^i E)=D(\mu^i, \lambda^i)=0.
$$
We assume that the matrix $A+\mu B+\lambda E$ has a simple spectrum for generic $\mu$ and $ \lambda$ satisfying the dispersion relation. 
 Then  relations (\ref{2}) give  expressions for the derivatives of $u^s, \ s\geq 2$, in terms of the derivatives of $u^1=u$,
\begin{equation}
\partial_1u^s=M^s(\mu^1, \lambda^1) \partial_1u, ~~~ \partial_2u^s=M^s(\mu^2, \lambda^2) \partial_2u, ~~~ s=2, ..., m,
\label{MN}
\end{equation}
where $M^s(\mu, \lambda)$  are rational functions of $\mu$ and $ \lambda$ whose coefficients are   determined by  $A$ and $B$. Applying the differentiation $\partial_j, \ j\ne i$, to the dispersion relation $D(\mu^i, \lambda^i)=0$, taking into account (\ref{MN}) and the commutativity conditions (\ref{comm}), one obtains
\begin{equation}
\frac{\partial_2\mu^1}{\mu^2-\mu^1}=\frac{\partial_2\lambda^1}{\lambda^2-\lambda^1}=P(\mu^1, \mu^2, \lambda^1, \lambda^2)\partial_2u, ~~~
\frac{\partial_1\mu^2}{\mu^1-\mu^2}=\frac{\partial_1\lambda^2}{\lambda^1-\lambda^2}=P(\mu^2, \mu^1, \lambda^2, \lambda^1)\partial_1u 
\label{PQ}
\end{equation}
where, again, $P$ is rational in its arguments. Finally, the consistency conditions
for the equations (\ref{MN}), $\partial_1\partial_2u^s=\partial_2\partial_1u^s$, 
 imply $m-1$ (a priory different) expressions for
$\partial_1\partial_2u$:
$$
\partial_1\partial_2u=R^s(\mu^1, \mu^2, \lambda^1, \lambda^2)\partial_1u \partial_2u.
$$
Thus, the condition of the existence of double waves is
$$
R^s(\mu^1, \mu^2, \lambda^1, \lambda^2)=R^l(\mu^1, \mu^2, \lambda^1, \lambda^2) ~~ {\rm mod} ~~
D(\mu^1, \lambda^1)=D(\mu^2, \lambda^2)=0.
$$
Since coefficients of the rational expressions $R^s,  R^l$  depend on the first derivatives of the matrix elements of $A$ and $B$, this  constitutes a set of nontrivial first order constraints for $A$ and $B$. 

An informal way to demonstrate the necessity part of Theorem 2 is the following: all of the above formulae possess the specialization  $\lambda^i=k+l\mu^i, ~ k,l =const$ (this is actually a subtle point which requires a justification). Then the dispersion relation reduces to $det (kE+A+\mu^i(lE+B))=0$, while the
relation (\ref{2}) takes the form $(kE+A+\mu^i(lE+B))\ \partial_i{\bf u}=0$. Thus,  $\partial_1{\bf u}$ and 
$\partial_2{\bf u}$ are the eigenvectors of  the matrix $(lE+B)^{-1}(kE+A)$ corresponding to the eigenvalues $\mu^1$ and $\mu^2$. Since the surface ${\bf u}(R^1, R^2)$ is tangential to both  eigenvectors, the two-dimensional distribution spanned by them is automatically holonomic. Since this is true for {\it any} two eigenvectors of the matrix $(lE+B)^{-1}(kE+A)$, it is  diagonalizable and, hence, has zero Haantjes tensor. A more rigorous proof of this result is given  below.

\medskip

\noindent {\bf Theorem 2}  {\it The Haantjes tensor of an arbitrary matrix (\ref{small}) is zero if and only if 

\noindent (i) the system (\ref{1})  possesses double waves
parametrized by four arbitrary functions of a single argument;

\noindent (ii) the characteristic speeds $\lambda^i, \ \mu^i$ of two-component reductions are not restricted by any  algebraic relations other than  the dispersion relation $D(\mu, \lambda)=0$. That is, for any point ${\bf u}_0$ in the hodograph space and any two points $(\lambda^1, \mu^1)$ and $(\lambda^2, \mu^2)$ on the dispersion curve at ${\bf u}_0$ one can find a two-component reduction whose characteristic speeds at this point take the values $(\lambda^1, \lambda^2)$ and $(\mu^1, \mu^2)$, respectively.}

\medskip

\centerline{\bf Proof:}

\medskip

Let  the system (\ref{1}) possess sufficiently many double waves satisfying the conditions (i), (ii) of the Theorem.  To prove the necessity part of Theorem 2 we need to show that the Haantjes tensor of any matrix (\ref{small}) equals zero. Let us fix  ${\bf u}_0$ in the hodograph space and consider a line $\lambda=k+l\mu$ in the $(\lambda, \mu)$-plane; here $k, l=const$ are the same as in (\ref{small}). The intersection of this line with the dispersion curve $D(\mu, \lambda)=0$ at ${\bf u}_0$ consists of $m$ points $(\lambda^1, \mu^1), ...,
(\lambda^m, \mu^m)$. Let us fix any two of them, say, $(\lambda^1, \mu^1)$ and $(\lambda^2, \mu^2)$. According to our assumptions there exists a double wave solution passing through ${\bf u}_0$  such that the characteristic speeds of the corresponding two-component reduction at the point  ${\bf u}_0$ take the values $(\lambda^1, \lambda^2)$ and $(\mu^1, \mu^2)$, respectively. We thus have
$$
(A+\mu^iB+\lambda^i E)\ \partial_i{\bf u}=0,
$$
$i=1, 2$. Setting $\lambda^i=K+L\mu^i$ (notice that at the point  ${\bf u}_0$ the values of $K$ and $L$ coincide with $k$ and $l$, respectively) we obtain
$$
(KE+A+\mu^i(LE+B))\ \partial_i{\bf u}=0.
$$
Introducing $V=-(LE+B)^{-1}(KE+A)$
we arrive at  $V\partial_i{\bf u}=\mu^i\partial_i{\bf u}$ or, using indices, 
\begin{equation}
V^m_n\partial_iu^n=\mu^i\partial_iu^m;
\label{tt}
\end{equation}
(no summation over $i$). Substituting $\lambda^i=K+L\mu^i$ into the commutativity conditions (\ref{comm}) we obtain the relations
\begin{equation}
\mu^i\partial_jL+\partial_j K=0, 
\label{ttt}
\end{equation}
$ i\ne j, \ i, j =1, 2$. 
Applying  to (\ref{tt}) the operator $\partial_j, \ j\ne i$, we obtain
\begin{equation}
V^m_{n,k}\partial_iu^n\partial_ju^k+V^m_n\partial_i\partial_j u^n=\partial_j\mu^i \partial_iu^m+\mu^i \partial_i\partial_j u^m.
\label{V}
\end{equation}
It is important to emphasize that the derivatives of $K$ and $L$ in the left hand side  cancel out by virtue of (\ref{ttt}). Thus,  both $K$ and $L$ behave like `constants'  in all tensor formulas below.
Interchanging   $i$ and $j$ in (\ref{V}) and subtracting the results  we arrive at the expression for 
$\partial_i\partial_j u^m$ in the form
\begin{equation}
\partial_i\partial_j u^m=\frac{\partial_j\mu^i}{\mu^j-\mu^i}\partial_iu^m
+\frac{\partial_i\mu^j}{\mu^i-\mu^j}\partial_ju^m
+\frac{V^m_{n,k}-V^m_{k,n}}{\mu^i-\mu^j}\partial_iu^n\partial_ju^k.
\label{2der}
\end{equation}
Substituting this back into (\ref{V}) we obtain a simple relation
$$
 \partial_j\mu^i \partial_iu^m+ \partial_i\mu^j \partial_ju^m=\frac{N^m_{nk} \partial_iu^n\partial_ju^k}{\mu^i-\mu^j}
 $$
 where $N$ is the Nijenhuis tensor of $V$  ($K$ and $ L$ can be regarded as `constants' in the computation of $N$).
 This  can be rewritten in the invariant form
 \begin{equation}
 \partial_j\mu^i \partial_i{\bf u}+ \partial_i\mu^j \partial_j{\bf u}=\frac{N(\partial_i{\bf u}, \ \partial_j{\bf u})}{\mu^i-\mu^j}
 \label{NN}
 \end{equation}
which implies the following four relations:
$$
\begin{array}{c}
(\mu^i)^2 \partial_j\mu^i \partial_i{\bf u}+ (\mu^j)^2\partial_i\mu^j \partial_j{\bf u}=\frac{V^2N(\partial_i{\bf u}, \ \partial_j{\bf u})}{\mu^i-\mu^j}, \\
\ \\
(\mu^i)^2 \partial_j\mu^i \partial_i{\bf u}+ \mu^i\mu^j\partial_i\mu^j \partial_j{\bf u}=\frac{VN(V\partial_i{\bf u}, \ \partial_j{\bf u})}{\mu^i-\mu^j}, \\
\ \\
\mu^i\mu^j \partial_j\mu^i \partial_i{\bf u}+ (\mu^j)^2\partial_i\mu^j \partial_j{\bf u}=\frac{VN(\partial_i{\bf u}, \ V\partial_j{\bf u})}{\mu^i-\mu^j}, \\
\ \\
\mu^i\mu^j \partial_j\mu^i \partial_i{\bf u}+ \mu^i \mu^j\partial_i\mu^j \partial_j{\bf u}=\frac{N(V\partial_i{\bf u}, \ V\partial_j{\bf u})}{\mu^i-\mu^j}.
\end{array}
$$
For instance, the first relation can be obtained by applying the operator $V^2$ to  (\ref{NN}) and 
using $V\partial_i{\bf u}=\mu^i\partial_i{\bf u}$, etc.  Taking a linear combination of the above relations we obtain
$$
V^2N(\partial_i{\bf u}, \ \partial_j{\bf u})-VN(V\partial_i{\bf u}, \ \partial_j{\bf u})-VN(\partial_i{\bf u}, \ V\partial_j{\bf u})+N(V\partial_i{\bf u}, \ V\partial_j{\bf u})=0.
$$
The last formula can be rewritten in the form $H(\partial_i{\bf u}, \ \partial_j{\bf u})=0$ where $H$ is the Haantjes tensor, indeed, a coordinate-free form of  (\ref{H}) is
$$
H(X, Y)=V^2N(X, Y)-VN(VX, Y)-VN(X,  VY)+N(VX,  VY)
$$
where $X, Y$ are arbitrary vector fields. Thus, the value of $H$  on any pair of eigenvectors of the matrix $V$ equals zero (the choice $i=1, \ j=2$ is not essential).  Hence, the Haantjes tensor of the matrix
$V=-(lE+B)^{-1}(kE+A)$ equals zero at the point ${\bf u}_0$, so that  the Haantjes tensor of   
$(kE+A)^{-1}(lE+B)$ is zero, too. Since ${\bf u}_0$ is arbitrary, this finishes the proof of the necessity part of Theorem 2. 

\medskip

The above considerations can be readily inverted to establish the sufficiency part of  the theorem. To construct double waves one needs to solve the equations (\ref{tt}), (\ref{ttt}). As demonstrated above, the consistency conditions for (\ref{tt}) reduce to 
(\ref{2der}) and (\ref{NN}), respectively. Among the second order equations (\ref{2der}) only one is really essential, say, the equation for the first component $u^1$ of the vector ${\bf u}$. The other are satisfied identically modulo (\ref{tt}) and (\ref{NN}). 
It remains to point out that the vanishing of the Haantjes tensor is equivalent to the condition that the relation (\ref{NN}) is an identity: first of all,  the vanishing of the Haantjes tensor implies that 
$N(\partial_i{\bf u}, \ \partial_j{\bf u})$ belongs to the span of $\partial_i{\bf u}$ and $ \ \partial_j{\bf u}$. 
Using the invariant definition of the Nijenhuis tensor
$$
N(X, Y)=[VX, VY]+V^2[X, Y]-V[X, VY]-V[VX, Y]
$$
one obtains that the coefficients  at $\partial_i{\bf u}$ and $ \ \partial_j{\bf u}$  in both sides of (\ref{NN}) 
coincide identically.
Thus, two-component reductions are governed by two first order relations (\ref{ttt}) and one second order equation for $u^1$. Up to reparametrizations $R^1\to f^1(R^1), \  R^2\to f^2(R^2)$ this leaves two arbitrary functions of one variable. Solving the corresponding hydrodynamic systems gives two extra arbitrary functions. Therefore, double waves depend on four arbitrary functions. 
Since $K$ and $L$ are non restricted by any algebraic relations, the condition (ii) is also satisfied.


\medskip

\section{A class of diagonalizable Godunov's systems}

In this section we discuss a class of conservative $3\times 3$ systems 
$$
{\bf u}_t+A({\bf u}) {\bf u}_x+B({\bf u}) {\bf u}_y=0
$$
where ${\bf u}=(u^1, u^2, u^3)^t$, $A$ is a constant diagonal matrix and $B$ is a Hessian:
$$
A=\left(\begin{array}{ccc}
a & 0 & 0\\
0 & b & 0\\
0 &0&c
\end{array}
\right), ~~~
B=\left(\begin{array}{ccc}
F_{11} & F_{12} & F_{13}\\
F_{21} & F_{22} & F_{23}\\
F_{31}&F_{32}&F_{33}
\end{array}
\right);
$$
here  $c<b<a$ are constants, $F_{ij}=\partial^2F/\partial u^i\partial u^j$. These systems belong to the class introduced in \cite{Godunov}. In the $2\times 2$ case the classification of integrable systems of this type was given in \cite{Fer5}. Here we concentrate on the diagonalizability aspect (which is automatically satisfied in the $2\times 2$ case). Calculating the    Haantjes tensor $H^i_{jk}$ for the matrix $(kE+A)^{-1}(lE+B)$ and equating it to zero one arrives at an over-determined system of third order PDEs for the potential $F$. The simplest way to obtain these equations is the following. One first calculates the components $H^i_{ij}$ (two of the indices coincide) and writes each of them with common denominator. Numerators thereof are polynomials in $k$ and $l$ which are required to be identically zero. Setting successively $k=-a, k=-b$ and $ k=-c$ in the expressions for these polynomials one arrives at the first set of relations,
\begin{equation}
\begin{array}{c}
F_{123}=0, \\
\ \\
F_{111}(b-c)=F_{122}(c-a)+F_{133}(a-b),\\
F_{222}(c-a)=F_{112}(b-c)+F_{332}(a-b), \\
F_{333}(a-b)=F_{113}(b-c)+F_{223}(c-a), \\
\ \\
F_{111}F_{23}=(F_{12}F_{13})_1, ~~
F_{222}F_{13}=(F_{21}F_{23})_2, ~~
F_{333}F_{12}=(F_{31}F_{32})_3, ~~ \\
\ \\
F_{111}=((c-a)(F_{12}^2)_1+(a-b)(F_{13}^2)_1)/{\triangle}, \\
F_{222}=((b-c)(F_{12}^2)_2+(a-b)(F_{23}^2)_2)/{\triangle}, \\
F_{333}=((b-c)(F_{13}^2)_3+(c-a)(F_{23}^2)_3)/{\triangle}, \\
\label{F}
\end{array}
\end{equation}
where $\triangle=(b-c)F_{11}+(c-a)F_{22}+(a-b)F_{33}$. Before proceeding we point out that the system (\ref{F}) possesses two integrals, one quadratic and one fourth order,
\begin{equation}
\begin{array}{c}
I=\triangle^2+4(c-a)(c-b)F_{12}^2+4(b-a)(b-c)F_{13}^2+4(a-b)(a-c)F_{23}^2,\\
\ \\
J=(c-b)F_{12}^2F_{13}^2+(b-a)F_{13}^2F_{23}^2+(a-c)F_{12}^2F_{23}^2
+F_{12}F_{13}F_{23}\triangle;
\end{array}
\label{IJ}
\end{equation}
both $I$ and $J$ are constant by virtue of (\ref{F}). The further analysis splits into two cases.

\noindent {\bf Case 1.} At least one of the expressions 
\begin{equation}
(a-c)F_{12}^2+(a-b)F_{13}^2, ~~
(b-a)F_{23}^2+(b-c)F_{21}^2, ~~ (c-a)F_{32}^2+(c-b)F_{31}^2
\label{C1}
\end{equation}
equals zero (these expressions appear as denominators when one solves the  system (\ref{F}) for the third derivatives of $F$). Let us suppose, for definiteness, that $(b-a)F_{23}^2+(b-c)F_{21}^2=0$ (notice that, since $c<b<a$, the other two possibilities lead to complex-valued solutions). Differentiating this 
constraint by $u^1$ and $u^3$ and taking into account $(\ref{F})_1$ one obtains   $F_{112}=F_{233}=0$. The substitution into $(\ref{F})_3$ implies $F_{222}=0$. The further integration of the system (\ref{F}) shows that, up to elementary transformations of variables, one has  the following expression for $F$:
$$
F=u^2(\gamma u^1+\delta u^3)+f(\gamma u^1+\delta u^3);
$$
here the constants $\gamma$ and $\delta$ satisfy the relation
$(b-a)\delta^2+(b-c)\gamma^2=0$ and $f$ is an arbitrary function of the indicated argument. 
We refer to the Example 2 of Sect. 2 for a detailed  discussion of this case.

\noindent {\bf Case 2.} All expressions (\ref{C1}) are nonzero. In this case one can solve the relations
(\ref{F}) for the third derivatives of $F$.  Calculating the remaining components $H^i_{jk}, \ i\ne j\ne k$, of the Haantjes tensor for the matrix $(kE+A)^{-1}(lE+B)$ and equating them to zero one arrives at the three cases:

\noindent 2a. All third order derivatives of $F$ equal to zero. This corresponds to linear systems. 

\noindent 2b. The integral $J$ equals zero. This case can be eliminated by the further consistency analysis. 

\noindent 2c. The integrals $I$ and $J$ satisfy the relation
\begin{equation}
I^2+64(a-b)(b-c)(c-a)J=0.
\label{relation}
\end{equation}
Notice that solutions constructed in Case 1 satisfy this constraint. We will show that there exists no other solutions in this class. First of all we  point out that the general (real-valued) solution to the first four equations (\ref{F}) (which are linear in $F$) is given by the formula
$$
F=g(z)+\bar g(\bar z)+p(\alpha) + q(z+\bar z+\alpha)+s(\alpha+z)+\bar s(\alpha+\bar z)+T
$$
where 
$$
z=\frac{u^1}{\sqrt{(b-c)}}+i\frac{u^2}{\sqrt{(a-c)}}, ~~
\bar z=\frac{u^1}{\sqrt{(b-c)}}-i\frac{u^2}{\sqrt{(a-c)}}, ~~
\alpha=-\frac{u^1}{\sqrt{(b-c)}}-\frac{u^3}{\sqrt{(a-b)}}
$$ 
and $T$ is an arbitrary quadratic form in $u^1, u^2, u^3$. Without any loss of generality one can assume $T= 2\epsilon \frac{(u^1)^2}{b-c}$. Substituting this ansatz for $F$ into (\ref{IJ}) and keeping in mind the relation (\ref{relation}) one arrives at the following set of functional equations for the functions 
$G=g'', ~ P=p'', ~ Q=q'', ~ S=s''$:
\begin{equation}
\begin{array}{c}
\left(G(z)+\bar G(\bar z)+P(\alpha)+Q(z+\bar z+\alpha)+S(\alpha+z)+\bar S(\alpha+\bar z)+2\epsilon \right) ^2- \\
\left(G(z)-\bar G(\bar z)\right) ^2-\left(P(\alpha)-Q(z+\bar z+\alpha)\right) ^2-\left(S(\alpha+z)-\bar S(\alpha+\bar z)\right) ^2={I}/{4}, 
\end{array}\label{F1}
\end{equation}
\begin{equation}
\begin{array}{c}
\left(G(z)-\bar G(\bar z)\right)^2 \left(P(\alpha)-Q(z+\bar z+\alpha)\right)^2+ \\
\left(G(z)-\bar G (\bar z)\right)^2 \left(S(\alpha+z)-\bar S(\alpha+\bar z)\right)^2+\\
\left(P(\alpha)-Q(z+\bar z+\alpha)\right)^2 \left(S(\alpha+z)-\bar S(\alpha+\bar z)\right)^2+ \\
2\left(G(z)-\bar G(\bar z)\right)\left(P(\alpha)-Q(z+\bar z+\alpha)\right)\left(S(\alpha+z)-\bar S(\alpha+\bar z)\right)\times \\
\left(G(z)+\bar G(\bar z)+P(\alpha)+Q(z+\bar z+\alpha)+S(\alpha+z)+\bar S(\alpha+\bar z)+2\epsilon \right)={I^2}/{64}.
\end{array}
\label{F2}
\end{equation}
These functional equations can be solved explicitly as follows.  Imposing the constraint $G(z)=\bar G(\bar z)$ in the second equation  (\ref{F2}) one obtains  
$\left(P(\alpha)-Q(z+\bar z+\alpha)\right)^2 \left(S(\alpha+z)-\bar S(\alpha+\bar z)\right)^2={I^2}/{64}$,
so that 
\begin{equation}
\left(P(\alpha)-Q(z+\bar z+\alpha)\right) \left(S(\alpha+z)-\bar S(\alpha+\bar z)\right)={I}/{8}
\label{I}
\end{equation}
(the case of the opposite sign is considered in a similar way).  Setting $G(z)=\bar G(\bar z)$ in the first equation  (\ref{F1}) we have
$$
\begin{array}{c}
\left(G(z)+\bar G(\bar z)+P(\alpha)+Q(z+\bar z+\alpha)+S(\alpha+z)+\bar S(\alpha+\bar z)+2\epsilon \right) ^2 =\\
\left(P(\alpha)-Q(z+\bar z+\alpha)\right) ^2+\left(S(\alpha+z)-\bar S(\alpha+\bar z)\right) ^2+{I}/{4}=\\
\left(P(\alpha)-Q(z+\bar z+\alpha)+S(\alpha+z)-\bar S(\alpha+\bar z)\right) ^2
\end{array}
$$
by virtue of (\ref{I}).
Thus, 
$$
\begin{array}{c}
G(z)+\bar G(\bar z)+P(\alpha)+Q(z+\bar z+\alpha)+S(\alpha+z)+\bar S(\alpha+\bar z)+2\epsilon=\\
-P(\alpha)+Q(z+\bar z+\alpha)-S(\alpha+z)+\bar S(\alpha+\bar z)
\end{array}
$$
(the case of the opposite sign is considered in a similar way) so that
$G(z)+P(\alpha)+S(\alpha+z)+\epsilon=0$. This simple functional equation implies that $G, \bar G, P, S, \bar S$ are linear functions:
$$
\begin{array}{c}
G(z)=cz+\mu, ~~ \bar G (\bar z)=c\bar z +\bar \mu, ~~ P(\alpha)=c\alpha +\nu, \\
S(\alpha+z)=-c(\alpha +z)+\eta, ~~ \bar S(\alpha +\bar z)=-c(\alpha +\bar z)+\bar \eta.
\end{array}
$$
However, the substitution of these expressions into (\ref{F1}) readily implies $c=0$. Thus, $G,  S $ and
$P$ are constants. This brings us back to the Case 1 considered above.

\section{Double waves for two-dimensional gas dynamics}

The equations of two-dimensional isentropic gas dynamics are of the form
$$
\rho_t+(\rho u)_x+(\rho v)_y=0, ~~ u_t+uu_x+vu_y+p_x/\rho=0, ~~ v_t+uv_x+vv_y+p_y/\rho=0,
$$
where $p=p(\rho)$ is the equation of state. In matrix form, one has ${\bf u}_t+A {\bf u}_x+B {\bf u}_y=0$ where ${\bf u}=(\rho, \ u,\  v)^t$ and
$$
A=\left(\begin{array}{ccc}
u & \rho & 0\\
c^2/\rho & u & 0\\
0&0&u
\end{array}
\right), ~~~
B=\left(\begin{array}{ccc}
v & 0 & \rho\\
0 & v & 0\\
c^2/\rho &0&v
\end{array}
\right);
$$
here $c^2=p'(\rho)$ is the sound speed. We have verified that, although the Haantjes tensor of an arbitrary matrix from the linear pencil $A+kB$ equals zero, this is not the case for the general family (\ref{small}). In particular, the Haantjes tensor of the matrix $A^{-1}B$ does not vanish. Nevertheless, equations of gas dynamics possess double waves of  special type, namely, potential double waves. These solutions have been extensively investigated in \cite{Sidorov}, see also references therein. We discuss them in a  different setting below. Let us first recall that the dispersion relation
$\det (E+\lambda A+\mu B)=0$ factorises into a line and a conic,
$$
(1+\lambda u+\mu v)((1+\lambda u+\mu v)^2-c^2(\lambda^2+\mu^2))=0,
$$
the conical branch corresponding  to sound waves. As demonstrated in \cite{Sidorov}, there exist  potential flows describing nonlinear interaction of two sound waves which are locally parametrized by four arbitrary functions of a single argument.  To construct these solutions we use  the ansatz
$\rho=\rho(R^1, R^2), \ u=u(R^1, R^2), \ v=v(R^1, R^2)$ where the Riemann invariants $R^1, R^2$ solve a pair of diagonal systems
\begin{equation}
R^i_x=\lambda^i({\bf R }) R^i_t, ~~~ R^i_y=\mu^i({\bf R }) R^i_t,
\label{dw}
\end{equation}
$ i=1, 2.$ The substitution implies the relations
$$
\begin{array}{c}
(1+\lambda^i u+\mu^i v)\partial_iu+\frac{c^2}{\rho}\lambda^i\partial_i\rho=0, \\
\ \\
(1+\lambda^i u+\mu^i v)\partial_iv+\frac{c^2}{\rho}\mu^i\partial_i \rho = 0, \\
\ \\
(1+\lambda^i u+\mu^i v)\partial_i\rho+\rho(\lambda^i\partial_i u + \mu^i\partial_iv)=0,
\end{array}
$$
$i=1, 2, \ \partial_i=\partial/\partial  R^i$. Expressing $\partial_i u$ and $\partial_i v$ from the first two equations and substituting them into the third, one arrives at  the dispersion relation for sound waves,
$$
(1+\lambda^i u+\mu^i v)^2-c^2((\lambda^i)^2+(\mu^i)^2) = 0.
$$
Parametrising $\lambda^i$ and $\mu^i$ in the form
$\lambda^i=s^i\cos \varphi^i, ~ \mu^i=s^i\sin \varphi^i, $
we obtain $1+\lambda^i u+\mu^i v=cs^i$ so that $s^i=1/(c-u\cos \varphi^i-v\sin \varphi^i)$. Thus,
\begin{equation}
\lambda^i=\frac{\cos \varphi^i}{c-u\cos \varphi^i-v\sin \varphi^i}, ~~~
\mu^i=\frac{\sin \varphi^i}{c-u\cos \varphi^i-v\sin \varphi^i}.
\label{skor}
\end{equation}
The equations for $u$ and $v$ take the form
\begin{equation}
\begin{array}{c}
\partial_1u+\frac{c}{\rho}\cos \varphi^1 \partial_1 \rho=0, ~~~ \partial_2u+\frac{c}{\rho}\cos \varphi^2 \partial_2 \rho=0, \\
 \  \\
 \partial_1v+\frac{c}{\rho}\sin \varphi^1 \partial_1 \rho=0, ~~~ \partial_2u+\frac{c}{\rho}\sin \varphi^2 \partial_2 \rho=0.
 \end{array}
 \label{uv}
 \end{equation}
 Notice that since $\mu^i \partial_iu=\lambda^i \partial_iv$ one has $u_y=v_x$. Thus, solutions describing nonlinear interaction of two sound waves are necessarily potential. Writing out the commutativity conditions 
 $\partial_j\lambda^i/(\lambda^j-\lambda^i)=\partial_j\mu^i/(\mu^j-\mu^i)$,  $i, j=1, 2, \ i\ne j$, and the consistency conditions $\partial_1\partial_2u=\partial_2\partial_1u, \  \partial_1\partial_2v=\partial_2\partial_1v$, one arrives at the following  system for $\varphi^1, \varphi^2, \rho$:
 \begin{equation}
 \begin{array}{c}
 \partial_2 \varphi^1=\cot {\frac{\varphi^2-\varphi^1}{2}}\left(\frac{c'}{c}+\frac{1}{\rho}\cos {(\varphi^1-\varphi^2)}\right) \partial_2 \rho, \\
\ \\
 \partial_1 \varphi^2=\cot {\frac{\varphi^1-\varphi^2}{2}}\left(\frac{c'}{c}+\frac{1}{\rho}\cos {(\varphi^1-\varphi^2)}\right) \partial_1 \rho, \\ 
 \ \\
 \partial_1\partial_2\rho=\displaystyle{\frac{\partial_1\rho \partial_2\rho}{\sin^2{\frac{\varphi^1-\varphi^2}{2}}}}
 \left(\frac{c'}{c}\cos {(\varphi^1-\varphi^2)} +\frac{1}{4\rho}(3+\cos {2(\varphi^1-\varphi^2)})
 \right).
 \end{array}
 \label{gas}
 \end{equation}
 Up to  reparametrizations $R^i\to f^i(R^i)$, the general solution of this system depends on two arbitrary functions of a single argument. For any solution $\varphi^i(R^1, R^2), \ \rho(R^1, R^2)$ one reconstructs the hodograph surface of the corresponding double wave by solving the equations (\ref{uv})  for  $u(R^1, R^2)$ and $v(R^1, R^2)$ (which are automatically consistent). Each of these  surfaces can be made into a solution of the gas dynamics equations by  solving the equations (\ref{dw}) for $R^1(t, x, y)$ and $R^2(t, x, y)$; here the characteristic speeds $\lambda^i, \mu^i$ are given by (\ref{skor}). Notice that since the general solution of (\ref{dw}) depends on two arbitrary functions of a single argument, each hodograph surface corresponds to infinitely many solutions. 
 
 \medskip 
 
 {\bf Remark.}  Any hodograph surface defined parametrically in the form $u(R^1, R^2), $ 
 $ v(R^1, R^2), $ $\rho(R^1, R^2)$ can be parametrized explicitly as $\rho=\rho(u, v)$. Using the relations (\ref{uv}), (\ref{gas}) one can show that the function $\rho(u, v)$  satisfies the second order PDE
\begin{equation}
 \begin{array}{c}
 \frac{1}{2}\left(\frac{\rho^2}{c^2}\right)'(\rho_u^2+\rho_v^2)-\frac{\rho^2}{c^2}
 \left(\frac{c'}{c}(\rho_u^2+\rho_v^2)+2\frac{\rho}{c^2}-(\rho_u^2+\rho_v^2)/\rho \right)= \\
 \ \\
 \frac{\rho^2}{c^2}(\rho_{uu}+\rho_{vv})-\rho_v^2\rho_{uu}+2\rho_u \rho_v \rho_{uv}-\rho_u^2\rho_{vv};
 \end{array}
 \label{rho}
 \end{equation}
 here $c^2=p'$ where $p(\rho)$ is the equation of state and $'\equiv d/d\rho$. For a polytropic gas, $p=\rho^{\gamma}/\gamma$, the equation (\ref{rho}) simplifies to 
 $$
(3-\gamma)\rho^{2-\gamma}(\rho_u^2+\rho_v^2)-2\rho^{5-2\gamma}=
\rho^{3-\gamma}(\rho_{uu}+\rho_{vv})-\rho_v^2\rho_{uu}+2\rho_u \rho_v \rho_{uv}-\rho_u^2\rho_{vv}
$$
which takes the form
$$
(1+\varphi_{uu})(\varphi_v^2-(\gamma-1)\varphi)-2\varphi_u\varphi_v\varphi_{uv}+
(1+\varphi_{vv})(\varphi_u^2-(\gamma-1)\varphi)=0
$$
after the substitution $\varphi =\rho^{\gamma-1}/(\gamma-1)$. In this form the equation for double waves appears in \cite{Sidorov}, \S 8.

 \medskip 
 
 It is unlikely that the system (\ref{gas}) can be integrated in a closed form, even for Chaplygin's equation of state $p(\rho)=b-a^2/\rho$, $a, b=const$, in which case it simplifies to
 $$
  \begin{array}{c}
 \partial_2 \varphi^1=\frac{ \partial_2 \rho}{\rho}\sin {(\varphi^1-\varphi^2)}, ~~~  \partial_1 \varphi^2=\frac{ \partial_1 \rho}{\rho}\sin {(\varphi^2-\varphi^1)}, \\ 
 \ \\
 \partial_1\partial_2\rho=2\frac{\partial_1\rho\partial_2\rho}{\rho}\sin^2{\frac{\varphi^1-\varphi^2}{2}}.
 \end{array}
$$
However, some particular solutions can be constructed. Let us consider the polytropic case, $p(\rho)=\rho^{\gamma}$, so that $\frac{c'}{c}=\frac{\gamma-1}{2\rho}$. One can  verify that 
$$
\varphi^1=R^1-R^2, ~~~ \varphi^2=R^1-R^2+\frac{\pi}{2}, ~~~ \rho=\exp^{-\frac{2}{\gamma-1}(R^1+R^2)}
$$
solve the  system (\ref{gas}). The corresponding equations (\ref{uv})  imply
$$
\begin{array}{c}
u=\frac{\sqrt {\gamma}}{1-\gamma } (\cos (R^1-R^2)-\sin (R^1-R^2))\exp^{-(R^1+R^2)}, \\
\ \\
v=\frac{\sqrt {\gamma}}{1-\gamma } (\cos (R^1-R^2)+\sin (R^1-R^2))\exp^{-(R^1+R^2)}.
\end{array}
$$
Excluding $R^1$ and $R^2$ one arrives at the explicit   hodograph surface, $\frac{(1-\gamma)^2}{2\gamma}(u^2+v^2)\rho^{1-\gamma}=1$, or, equivalently,
$(1-\gamma)^2(u^2+v^2)=2c^2$. To make this surface into a solution one has to solve the equations (\ref{dw}) where the characteristic speeds are given explicitly by (\ref{skor}):
$$
\begin{array}{c}
\lambda^1=\frac{\gamma-1}{\gamma\sqrt {\gamma}}\cos (R^1-R^2)\exp ^{R^1+R^2}, ~~~
\lambda^2=-\frac{\gamma-1}{\gamma\sqrt {\gamma}}\sin (R^1-R^2)\exp ^{R^1+R^2}, \\
\ \\
\mu^1=\frac{\gamma-1}{\gamma\sqrt {\gamma}}\sin (R^1-R^2)\exp ^{R^1+R^2}, ~~~
\mu^2=\frac{\gamma-1}{\gamma\sqrt {\gamma}}\cos (R^1-R^2)\exp ^{R^1+R^2}, 
\end{array}
$$
The general solution $R^1(t, x, y), \ R^2(t, x, y)$ of the corresponding system (\ref{dw}) is given by the implicit formula
$$
v^1({\bf R })=t+\lambda^1({\bf R })x+\mu^1({\bf R })y, ~~~ v^2({\bf R })=t+\lambda^2({\bf R })x+\mu^2({\bf R })y
$$
where $v^1, v^2$ are the characteristic speeds of the general flow commuting with (\ref{skor}). They satisfy the linear equations
$$
\begin{array}{c}
\frac{\partial_2v^1}{v^2-v^1}=\frac{\partial_2\lambda^1}{\lambda^2-\lambda^1}=\frac{\partial_2\mu^1}{\mu^2-\mu^1}=-1\\
\ \\
\frac{\partial_1v^2}{v^1-v^2}=\frac{\partial_1\lambda^2}{\lambda^1-\lambda^2}=\frac{\partial_1\mu^2}{\mu^1-\mu^2}=-1.
\end{array}
$$
Potential double waves of this  type and their applications to problems of gas dynamics were extensively investigated in the monograph \cite{Sidorov}. The aim of this section was to demonstrate how solutions with a degenerate hodograph fit into the general scheme of the method of hydrodynamic reductions proposed  in \cite{Fer4}. In particular, our derivation of equations governing potential double waves is, in our opinion, much more transparent and straightforward since the use of Riemann invariants $R^1, R^2$ considerably simplifies the calculation of compatibility conditions. On the contrary, the approach of \cite{Sidorov} leads to quite cumbersome calculations based on  Cartan's theory of over-determined systems.

\section{Appendix:  the semi-Hamiltonian property}

For diagonalizable one-dimensional systems (\ref{1d}) there exists a  tensor object responsible for the semi-Hamiltonian property. We emphasize that the method does {\it not} require the actual transformation  to the diagonal form (\ref{1R}): it is sufficient to verify that the corresponding Haantjes tensor (\ref{H}) is zero. After that one  computes the $(1, 3)$-tensors $M$ and $K$,
$$
M^s_{kij}=N^s_{kp}v^p_qN^q_{ij}+N^s_{pq}v^p_kN^q_{ij}-N^s_{pq}N^p_{ik}v^q_j
-N^s_{pq}N^p_{kj}v^q_i-N^s_{kp}N^p_{iq}v^q_j-N^s_{kp}N^p_{qj}v^q_i
$$
and
$$
\begin{array}{c}
K^s_{kij}=b^s_p\partial_{u^k}N^p_{ij}-b^p_k\partial_{u^p}N^s_{ij}+N^p_{ij}\partial_{u^p}b^s_k-
N^s_{kp}\partial_{u^i}b^p_j+N^s_{kp}\partial_{u^j}b^p_i \\
\ \\
+b^s_p\partial_{u^i}N^p_{jk}-b^p_i\partial_{u^p}N^s_{jk}+N^p_{jk}\partial_{u^p}b^s_i-
N^s_{ip}\partial_{u^j}b^p_k+N^s_{ip}\partial_{u^k}b^p_j \\
\ \\
+b^s_p\partial_{u^j}N^p_{ki}-b^p_j\partial_{u^p}N^s_{ki}+N^p_{ki}\partial_{u^p}b^s_j-
N^s_{jp}\partial_{u^k}b^p_i+N^s_{jp}\partial_{u^i}b^p_k;
\end{array}
$$
here $b=v^2$, that is, $b^i_j=v^i_pv^p_j$. Using $M$ and $K$ one defines the $(1, 3)$-tensor $Q$ as
$$
\begin{array}{c}
Q^s_{kij}=v^p_kK^s_{pqj}v^q_i+v^p_kK^s_{piq}v^q_j-v^p_qv^q_kK^s_{pij}-K^s_{kpq}v^p_iv^q_j\\
\ \\
+4v^p_kM^s_{pij}-2M^s_{kpj}v^p_i-2M^s_{kip}v^p_j.
\end{array}
$$
Finally, one introduces the tensor $P$,
\begin{equation}
P^s_{kij}=v^s_pQ^p_{kqj}v^q_i+v^s_pQ^p_{kiq}v^q_j-
v^s_qv^q_pQ^p_{kij}-Q^s_{kpq}v^p_iv^q_j.
\label{P}
\end{equation}
\begin{theorem} \cite{Pavlov2}
 A  diagonalizable hydrodynamic type system (\ref{1d}) with the matrix $v^i_j(u)$
with mutually distinct eigenvalues is semi-Hamiltonian if and only if the corresponding  tensor $P$ is identically zero. 
\end{theorem}
Note that these objects can be  obtained using computer algebra.

\section*{Acknowledgements}

We thank M.V. Pavlov and S.P. Tsarev for numerous discussions. We also thank the referees  for a constructive criticism and useful suggestions.


\begin{thebibliography}{99}


\bibitem{Bla} M. Blaszak, B.M. Szablikowski, Classical R-matrix
theory of dispersionless systems: II. (2+1)-dimension theory, J. Phys. A {\bf 35} (2002) 10345-10364.


\bibitem{Burnat1} M. Burnat, The method of Riemann invariants for
multi-dimensional nonelliptic system, Bull. Acad. Polon. Sci. Sr.
Sci. Tech. {\bf 17} (1969) 1019-1026.

\bibitem{Burnat2} M. Burnat, The method of Riemann invariants and
its applications to the theory of plasticity. I, II. Arch. Mech.
(Arch. Mech. Stos.) {\bf 23} (1971), 817-838; ibid. {\bf 24} (1972),
3-26.

\bibitem{Burnat3} M. Burnat, The method of characteristics and
Riemann's invariants for multidimensional hyperbolic systems,
 Sibirsk. Mat.  Z. {\bf 11} (1970) 279-309.

\bibitem{K} S. Canic and B. L. Keyfitz, A useful class of two-dimensional conservation laws,  Proceedings of ICIAM 95: Supplement 2: Applied Analysis, Mathematical Research, {\bf 87}, (K. Kirchgassner, O. Mahrenholtz and R. Mennicken, editors) Akademie Verlag, Berlin, ZAMM (1996) 133-136.

\bibitem{Dafermos} C. Dafermos, Hyperbolic conservation laws in
continuum physics,
Springer-Verlag, 2000.

\bibitem{Dinu} L. Dinu, Some remarks concerning the Riemann
invariance, Burnat-Peradzy\'nski
and Martin approaches, Rev. Roumaine Math. Pures Appl. {\bf 35}, N3
(1990) 203-234.

\bibitem{Dub} B.A. Dubrovin and S.P. Novikov, Hydrodynamics of weakly
deformed soliton lattices:
differential geometry and Hamiltonian theory, Russian Math. Surveys
{\bf 44} , N6 (1989) 35-124.

\bibitem{Dubr} B.A. Dubrovin, Geometry of 2D topological field
theories, Lect. Notes in Math.
{\bf 1620}, Springer-Verlag (1996) 120-348.


\bibitem{Fer} E.V. Ferapontov,  D. A. Korotkin  and V.A. Shramchenko,
Boyer-Finley equation and systems of hydrodynamic type, Class.
Quantum Grav. {\bf 19} (2002) L205-L210.

\bibitem{Fer2} E.V. Ferapontov and M.V. Pavlov, Hydrodynamic
reductions of the heavenly equation, Class. Quantum Grav. {\bf 20}
(2003) 2429-2441.

\bibitem{Fer3} E.V. Ferapontov and S.P. Tsarev, Systems of hydrodynamic type that arise in gas chromatography. Riemann invariants and exact solutions,  Mat. Model. {\bf 3} , N2 (1991) 82-91.

\bibitem{Fer4} E.V. Ferapontov and K.R. Khusnutdinova, On integrability of (2+1)-dimensional quasilinear systems, Comm. Math. Phys. {\bf 248} (2004) 187-206; arXiv:nlin.SI/0305044.

\bibitem{Fer5} E.V. Ferapontov and K.R. Khusnutdinova, The characterization of two-component (2+1)-dimensional integrable  systems
of hydrodynamic type,   J. Phys. A: Math. Gen. {\bf 37}, N8 (2004)
2949 - 2963; arXiv:nlin.SI/0310021.

\bibitem{Fer6} E.V. Ferapontov and K.R. Khusnutdinova, Hydrodynamic reductions of multi-dimensional dispersionless PDEs: the test for integrability,   J. Math. Phys. {\bf 45}, N6 (2004) 2365-2377;  arXiv:nlin.SI/0312015.

\bibitem{Fer7} E.V. Ferapontov and D.G. Marshall, Differential-geometric approach to the integrability of hydrodynamic chains: the Haantjes tensor,  arXiv:nlin.SI/0505013.


\bibitem{Gibb94} J. Gibbons and Y. Kodama,   A method for solving the
dispersionless KP hierarchy and its exact solutions. II. Phys. Lett.
A {\bf 135}, N3 (1989) 167--170.

\bibitem{GibTsa96} J. Gibbons and S. P. Tsarev, Reductions of the
Benney equations, Phys. Lett. A {\bf 211} (1996) 19-24.

\bibitem{GibTsa99} J. Gibbons and S. P. Tsarev, Conformal maps and
reductions of the Benney equations, Phys. Lett. A {\bf 258} (1999)
263-271.

\bibitem{Godunov} S.K. Godunov,  An interesting class of quasi-linear
systems,  Dokl. Akad. Nauk SSSR {\bf 139} (1961) 521-523.

\bibitem{Grundland} A. Grundland and R. Zelazny, Simple waves in
quasilinear hyperbolic systems.
   I, II. Riemann invariants for the problem of simple wave
interactions. J. Math. Phys. {\bf 24},
    N9 (1983)  2305-2328.

\bibitem{GuMaAl}F. Guil, M. Manas and L. Martinez  Alonso, On the
Whitham Hierarchies: Reductions and Hodograph Solutions, J. Phys. A {\bf 36} (2003) 4047-4062.

\bibitem{Haantjes} J.Ê Haantjes, On $X\sb m$-forming sets of eigenvectors, Indagationes Mathematicae {\bf 17}  (1955) 158-162.

\bibitem{Kr1} I.M.Ê Krichever, The averaging method for two-dimensional "integrable" equations, Funct. Anal. Appl. {\bf 22}, N3 (1988) 200-213.


\bibitem{Kr2} I.M.  Krichever,  Spectral theory of two-dimensional periodic operators and its applications, Russian Math. Surveys {\bf 44}, N2 (1989) 145-225.

\bibitem{Kr3} I.M. Krichever, M. Mineev-Weinstein, P. Wiegmann and A. Zabrodin, Laplacian growth and Whitham equations of soliton theory,  Phys. D {\bf 198}, no. 1-2 (2004) 1-28.

\bibitem{Majda} A. Majda, Compressible fluid flows and systems of
conservation laws in several space variables, Appl. Math. Sci.,
Springer-Verlag, NY, {\bf 53} (1984) 159 pp.


\bibitem{Ma} M. Manas,  L. Martinez  Alonso and E. Medina,
Reductions and hodograph solutions of the dispersionless KP
hierarchy, J. Phys. A: Math. Gen. {\bf 35} (2002) 401-417.

\bibitem{Ma1} M. Manas, $S$-functions, reductions and hodograph solutions of the $r$-th dispersionless modified KP and Dym hierarchies, J. Phys. A {\bf 37}, no. 46 (2004) 11191-11221.

\bibitem{Shabat} L. Martinez Alonso and A.B. Shabat, Hydrodynamic reductions and solutions of a universal hierarchy, Teoret. Mat. Fiz. {\bf 140}, no. 2 (2004) 216-229. 


\bibitem{Pavlov} M.V. Pavlov,  Integrable hydrodynamic chains, J. Math. Phys. {\bf 44} (2003) 4134-4156.

\bibitem{Pavlov1} M.V. Pavlov, Classification of the integrable
Egorov hydrodynamic chains, Theor. and Math. Phys.
 {\bf 138}, no 1 (2004) 55-70.


\bibitem{Pavlov2} M.V. Pavlov, S.I.  Svinolupov and R.A.  Sharipov,  An invariant criterion for hydrodynamic integrability, Funktsional. Anal. i Prilozhen. {\bf 30} (1996) 18-29; translation in Funct. Anal. Appl. {\bf 30} (1996) 15-22.

\bibitem{Perad1} Z. Peradzy\'nski, Riemann invariants for the
nonplanar $k$-waves, Bull. Acad. Polon. Sci. Sr. Sci. Tech. {\bf 19}
(1971) 717-724.

\bibitem{Perad2} Z. Peradzy\'nski, Nonlinear plane $k$-waves and
Riemann invariants, Bull. Acad. Polon. Sci. Sr. Sci. Tech. {\bf 19}
(1971) 625-632.


\bibitem{Sevennec} B.~S\'evennec, G\'eom\'etrie des syst\`emes hyperboliques
de lois de conservation, M\'emoire (nouvelle s\'erie) N56,
Suppl\'ement au Bulletin de la Soci\'et\'e Math\'ematique de France {\bf 122} (1994) 1-125.

\bibitem{Serre1} D. Serre, Systems of conservation laws. 1. 
Hyperbolicity, entropies, shock waves, Cambridge University Press
(1999) 263 pp.

\bibitem{Serre2} D. Serre, Systems of conservation laws. 2. 
Geometric structures, oscillations, and initial-boundary value problems, 
Cambridge University Press (2000) 269 pp. 

\bibitem{Sidorov} A.F. Sidorov, V.P. Shapeev and N.N. Yanenko, The method of differential constraints and its applications in gas dynamics, ``Nauka'', Novosibirsk (1984) 272 pp.

\bibitem{Tsarev} S.P. Tsarev, Geometry of hamiltonian systems of
hydrodynamic type. Generalized hodograph method, Izvestija AN USSR
Math. {\bf 54}, N5  (1990) 1048-1068.

\bibitem{Tsarev1} S.P. Tsarev, Diagonalizable averaging of the generalized KdV equation,
 Uspekhi Matem. Nauk.  {\bf 46}, N6 (1991) 194.

\bibitem{Lei}  L. Yu, Reductions of dispersionless integrable hierarchies,
PhD Thesis, Imperial College, London, 2001.


\bibitem{Zakharov}  E.V. Zakharov,  Dispersionless limit of
integrable systems in $2+1$ dimensions, in Singular Limits of
Dispersive Waves, Ed. N.M. Ercolani et al., Plenum Press, NY, (1994)
165-174.


\end{thebibliography}
\end{document}